\documentclass[doublecol,figures]{epl2} 
\usepackage[apple mac]{inputenc}
\usepackage{graphicx}
\usepackage{epstopdf}
\usepackage{picins}
\usepackage{epsf}
\usepackage{subfigure}
\usepackage{amsmath}
\usepackage{amssymb}
\usepackage{amsfonts}

\title{ Spatial patterns in optical lattices submitted to gauge potentials }

\author{ N. Goldman}
\institute{ Center for Nonlinear Phenomena and Complex Systems -
Universit\'e Libre de Bruxelles (U.L.B.), Code Postal 231, Campus Plaine, B-1050
Brussels, Belgium}


\abstract{We study the vortex formation in optical lattices submitted to artificial gauge potentials. We compute the superfluid density for Abelian and non-Abelian gauge potentials with a mean-field approach of the Bose-Hubbard model and we determine the rule describing the number of vortices as a function of the effective magnetic flux. This simple rule is represented by a remarkably rich figure that represents the superfluid density as a function of the flux. The phenomena which emanate from this work should be observed experimentally in optical lattices within which atom tunneling is laser-assisted and described by commutative or non-commutative tunneling operators.
} 

\begin{document}

\maketitle

The observation and comprehension of spatial patterns in physical systems constitute a major field of investigation\cite{Nicolis}. The study of these complex structures, which arise at the macroscopic level, is necessary in order to understand the behaviour of numerous quantum systems such as type-II superconductors \cite{vc,vc2,vc3,vc4}, rotating helium superfluids \cite{vl,vl2}, rotating Bose-Einstein condensates \cite{vb,vb2} and the fractional quantum Hall effect\cite{Gaspard2003}. 

In the context of superfluid Bose gases\cite{vb,vb2}, these observable patterns result from topological defects in the macroscopic wave function which constitute an order parameter for the insulator-superfluid transition\cite{Sheshadri,Greiner1}. These singular points in which the superfluid density dramatically decreases, known as vortices, are regarded as interacting entities and usually form regular lattices.

The very recent progresses in cooling and manipulation of cold atomic systems 
\cite{Chin2006,Kohl2005 ,Gunter2005,Stoferle2006} have open the way to investigations  of fundamental aspects of condensed matter physics. Optical lattices, created by adjusted lasers, constitute an elegant way  to study periodically trapped atomic systems\cite{Lewenstein,Jaksch}. Among the various experimental configurations that can now be set up is the application of artificial gauge potentials (artificial magnetic fields) to neutral atoms \cite{Zoller2003,Lewenstein,Sorensen2005,Osterloh2005,Ruseckas2005}. Such setups notably offer the possibility to reproduce the dynamics of periodically constrained electrons submitted to a magnetic field\cite{Zoller2003,Lewenstein}. This should lead to the observation of the corresponding fractal energy spectrum (the Hofstadter butterfly\cite{Hofstadter1976})  and to the exploration of the quantum Hall effects\cite{Sorensen2005,Goldman2007}. In particular the realization of non-Abelian gauge potentials has been envisaged\cite{Lewenstein}, allowing the observation of a non-Abelian Aharonov-Bohm effect \cite{Osterloh2005}, magnetic monopoles \cite{Ruseckas2005}, particular metal-insulator transitions \cite{Satija2006} and an integer quantum Hall effect for neutral currents\cite{Goldman2007}.  

In this context, a problem of interest is the effect of gauge potentials on the Mott insulator transition: It is known that the insulator-superfluid transition is affected by an Abelian gauge potential\cite{Oktel}: the Mott lobes, which characterize the corresponding phase diagram, are increased when the artificial magnetic field is turned on. Moreover the critical point of the transition is proportional to the bandwidth of the Hofstadter butterfly \cite{Hofstadter1976}, the energy spectrum of the noninteracting problem. 

In this work we are interested by the effects induced by a gauge potential once the system has reached the superfluid regime. Since the creation of an effective magnetic field within an optical lattice reproduces the dynamics of a spinning Bose system, as expected we observed in this work vortex configurations \cite{vb,vb2,Oktel,Wu,Burkov} that we studied for general non-Abelian gauge potentials. In this letter we first introduce the Bose-Hubbard Hamiltonian describing the optical lattice submitted to non-Abelian gauge potentials and we briefly describe the mean-field approach considered in this context. The vortex configurations are studied for Abelian and non-Abelian configurations of the external gauge potential. The general rule describing the pattern formation that emanates from this work is eventually exposed.

To start with, we consider a bosonic system trapped in a two-dimensional optical lattice and submitted to artificial gauge potentials. In this letter we consider a family of non-Abelian gauge potentials, which contains Abelian potentials as particular cases. The system is then described by a generalized Bose-Hubbard Hamiltonian written as a sum of single-site terms $\mathcal{H}=\sum_{n,m} \mathcal{H}_{n,m}$,
with
\begin{align} 
\mathcal{H}_{n,m}&= - t_a   \{ a^{\dagger}_{n,m} \vert U_x
\vert a_{n+1,m} \} - t_b    \{ a^{\dagger}_{n,m} \vert U_y \vert
a_{n,m+1} \}  \notag \\ 
& - t_a   \{ a^{\dagger}_{n,m} \vert U_x ^{\dagger}
\vert a_{n-1,m} \} - t_b    \{ a^{\dagger}_{n,m} \vert U_y ^{\dagger} \vert
a_{n,m-1} \}  \notag \\ 
&+ \frac{V}{2} \bigl (  \{ a^{\dagger}_{n,m} 
\vert a_{n,m} \} -1 \bigr ) \{ a^{\dagger}_{n,m} 
\vert a_{n,m} \} - \mu \{ a^{\dagger}_{n,m} 
\vert a_{n,m} \} \notag \\
\label{ham}
\end{align}
The first two lines of the expression determine the dynamics of  noninteracting ``charged bosons" in a two-dimensional
lattice within a tight-binding approximation as studied
experimentally in cold atom systems trapped in optical lattices\cite{Chin2006,Kohl2005,Gunter2005,Stoferle2006}. For
$U_{x,y}$ belonging to the Abelian group of unitary complex numbers $U(1)$, this noninteracting part of the Hamiltonian reproduces the evolution of an
electronic system constrained by a periodic potential and submitted
to a magnetic field \cite{Zoller2003}. In the present context one consider a
system with $U(2)$ gauge structure, which implies that the operators
$U_x$ and $U_y$ are $2 \times 2$ unitary matrices and the single-particle wave function has two components: when tunneling occurs along the $x$ ($y$) direction, the matrix $U_x$ ($U_y$) acts as a tunneling operator on the two-component wave function and the system is non-Abelian for $[U_x,U_y] \ne 0$ \cite{Osterloh2005,Satija2006}. The operators $U_{x,y}$
are related to the gauge potential $\boldsymbol{A}$ present in the system via the usual relation $U_{x,y}=e^{i A_{x,y}}$.
In order to indicate multicomponent  wave functions and operators, we use the following notations
introduced in Refs. \cite{Mead1992,Goldman2007}: A \emph{row} of $2$ orthogonal kets
 is denoted by $\{ \vert \psi \rangle \vert=(\vert \psi
\rangle_1,  \vert \psi \rangle _2)$ and a \emph{column}
of $2$ kets by $\vert \vert \psi \rangle
\}$.The coefficients
$t_a$ and $t_b$ describe transfers in the $x=n l_x$ and $y=m l_y$
directions, where $(l_x, l_y)$ are the unit cell's lengths, and $\vert a^{\dagger}_{n,m} \}$ (resp. $\vert a_{n,m}
\}$) are the $2$-component bosonic creation (resp. annihilation)
field operators on site $(n,m)$. The second line of the expression includes the interaction term and the chemical potential. 

The system is solved using a mean-field approach based on the decoupling formula\cite{Sheshadri,Oktel,Wu} 
\begin{align}
a_{n,m,j}^{\dagger} a_{n,m,j'}&=\langle a_{n,m,j}^{\dagger} \rangle a_{n,m,j'} +a_{n,m,j}^{\dagger} \langle a_{n,m,j'} \rangle \notag \\ 
&-\langle a_{n,m,j}^{\dagger} \rangle \langle a_{n,m,j'} \rangle \label{approx}
\end{align}
with $j,j'=1,2$. In the Abelian case, the average value $\langle a_{n,m,1} \rangle=\Psi _{n,m,1}=\Psi _{n,m,2}$ defines an order parameter for the insulator-superfluid transition and is non-zero when the ground state is in the superfluid state\cite{Oktel}. As we are dealing with a non-Abelian system for which $\Psi _{n,m,1} $ generally differs from $\Psi _{n,m,2}$ we define $\vert \Psi_{n,m} \vert ^2= \{ \Psi_{n,m} \vert \Psi_{n,m} \}$ as the superfluid density, and refer $\vert \Psi_{n,m,j} \vert ^2$ as a ``color" density with $j=1,2$.

We and others have recently studied the effects induced by non-Abelian gauge potentials in optical lattices \cite{Osterloh2005,Satija2006, Goldman2007}. Each vertex of a $3D$ optical
lattice contains an atom with degenerate Zeeman sublevels in the
hyperfine ground state manifolds $\{\vert g \rangle _j , \vert e
\rangle _j \}$ with $j=1,2$, so that the states we are dealing with are
represented by the rows $\{\vert e \rangle \vert=(\vert e \rangle _1,
\vert e \rangle _2)$ or $\{ \vert g \rangle \vert=(\vert g \rangle
_1, \vert g \rangle _2)$. Lasers are used in order to create the
lattice, as well as the non-Abelian gauge potential through
state-dependent  control of hoppings which take place within every
plane $z=$ constant. In such setups, atoms hop from a site to another and their tunnelings are described by unitary  non-commutative operators \cite{Osterloh2005}.  The corresponding gauge potential is given by
 \begin{equation}
\boldsymbol{A}=  \Biggl [\begin{pmatrix} -\frac{\pi}{2}
&\frac{\pi}{2} \\
\frac{\pi}{2} &-\frac{\pi}{2}
\end{pmatrix},\begin{pmatrix} 2 \pi n \alpha _1 &0 \\
0 &2 \pi n \alpha _2 \end{pmatrix},0 \Biggr]
\label{potential} \end{equation} and induces an effective
``magnetic" field characterized by two effective magnetic fluxes $\alpha _1$ and $\alpha _2$,  which can be controlled with external lasers \cite{Osterloh2005}.   This gauge potential is Abelian under the condition that $\alpha_1-\alpha_2$ is an integer and non-Abelian otherwise. The gauge potential preserves the translational symmetry along the $y$-axis, such that  $\vert \Psi _{n,m} \}=\vert \Psi_n \}$. We consider quantized fluxes $\alpha_1=p_1/q$ and $\alpha _2=p_2/q$, with $p_1, p_2, q$ integers, such that $\vert \Psi _{n+q} \}= \vert \Psi_n \}$. Including the potential $\boldsymbol{A}$ in the mean-field Hamiltonian (\ref{ham},\ref{approx}), one finds the non-zero matrix elements of $\mathcal{H}^{MF}_{n}$ in the occupation number basis $\vert N_{n,j} \rangle$, with $N_{n,j}=0,...,\infty$:

\begin{gather}
\begin{split}
&\langle N_{n,1} \vert \mathcal{H}^{MF}_{n} \vert N_{n,1} \rangle = \frac{V}{2} (N_{n,1}-1) N_{n,1}- \mu N_{n,1} \\
&+ t_a \bigl ( \Psi_{n,1}^* \Psi_{n+1,2} +  \Psi_{n,2}^* \Psi_{n+1,1} + h.c. \bigr ) \\
&+ t_b \bigl (2 \, cos (2 \pi \alpha_1 n)  \Psi_{n,1}^* \Psi_{n,1} + 2 \,  cos (2 \pi \alpha_2 n)  \Psi_{n,2}^* \Psi_{n,2} \bigr ) \notag
\end{split} \\
=\langle N_{n,2} \vert \mathcal{H}^{MF}_{n} \vert N_{n,2} \rangle \notag
\\
\begin{split}
\langle N_{n,1}+1 \vert \mathcal{H}^{MF}_{n} \vert N_{n,1} \rangle &= 
- t_a \sqrt{N_{n,1}+1} \bigl ( \Psi_{n+1,2} + \Psi_{n-1,2}  \bigr ) \\
&- 2 t_b \sqrt{N_{n,1}+1} \, cos (2 \pi \alpha_1 n)  \Psi_{n,1}  \notag
\end{split}
\\
\begin{split}
\langle N_{n,2}+1 \vert \mathcal{H}^{MF}_{n} \vert N_{n,2} \rangle &= 
- t_a \sqrt{N_{n,2}+1} \bigl ( \Psi_{n+1,1} + \Psi_{n-1,1}  \bigr ) \\
&- 2 t_b \sqrt{N_{n,2}+1} \, cos (2 \pi \alpha_2 n)  \Psi_{n,2}  
\label{mat}
\end{split} 
\end{gather}
where the expressions $a_{n,j} \vert N_{n,j}+1 \rangle = \sqrt{ N_{n,j}+1} \vert N_{n,j} \rangle$ and $a_{n,j}^{\dagger} \vert N_{n,j} \rangle = \sqrt{ N_{n,j}+1} \vert N_{n,j} +1 \rangle$ have been used. The ground state $\vert G_{n} \rangle$ is computed in a truncated basis, $N_{n,j}=1,..., N_{max}$, and the order parameter $\vert \Psi_n \}$ is given by $\Psi _{n,j} = \langle G_n \vert a_{n,j} \vert G_n \rangle$. The latter expressions  form a set of self-consistent equations which is solved in order to find the ground state $\vert G_n \rangle$ and the order parameter $\vert \Psi_n \}$.

\begin{figure}
\includegraphics[scale=1.1]{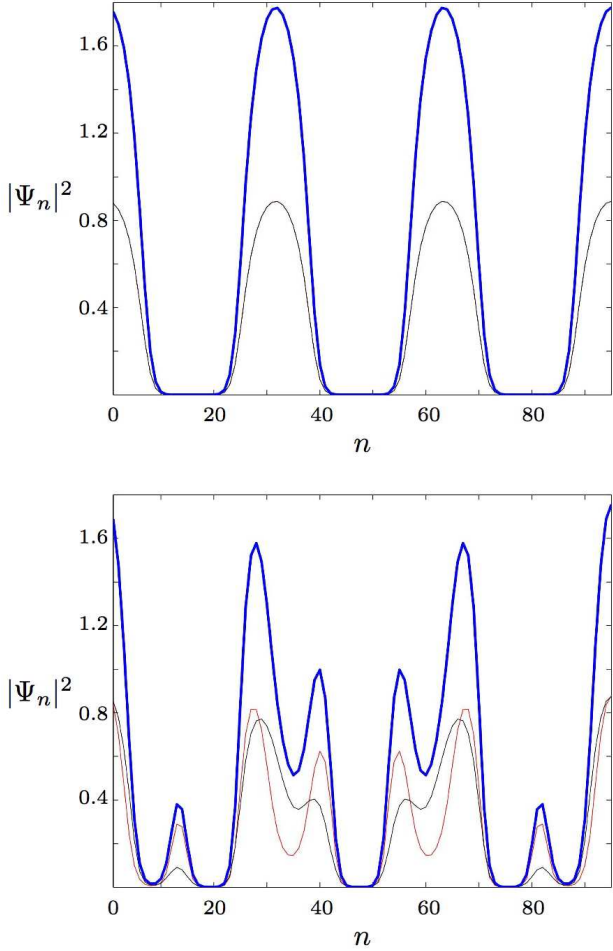}  \\
\caption{\label{plot} ``Color" densities $\vert \Psi_{n,1} \vert ^2$ (red line) and $ \vert \Psi_{n,2} \vert ^2$ (black line) and  superfluid density $\vert \Psi_{n} \vert ^2=\{ \Psi_n \vert \Psi_n \}$ (thick blue line) for $\mu=1.4$, $t_a=t_b=0.05$ and $q=95$: (a) $\alpha
_1= \alpha_2=\frac{3}{q}$;  (b) $\alpha
_1=\frac{7}{q}$ and $\alpha_2=\frac{3}{q}$ }
 \end{figure}

The system presents symmetries under translations along the $y$-axis and the system is $q$-periodic along the $x$-axis when the gauge potential \eqref{potential} is turned on: one compute the densities on a supercell $q \times 1$ with periodic boundary conditions. The superfluid density and the ``color" densities are depicted in Fig.\ref{plot} for specific Abelian (Fig.\ref{plot}.a) and non-Abelian (Fig.\ref{plot}.b)  situations when the system is in the superfluid regime ($\mu=1.4$, $t_a=t_b=0.05$). For both cases oscillations in the superfluid densities are observed; in the following we consider the local minima as \emph{vortices}.  In Fig.\ref{plot}, one set the supercell length to $q=95$ and one finds $3$ vortices when $\alpha
_1= \alpha_2=\frac{3}{q}$ (Fig.\ref{plot}.a). In the non-Abelian case (Fig.\ref{plot}.b) one finds $7$ vortices when $\alpha
_1=\frac{7}{q}$ and $\alpha_2=\frac{3}{q}$. We notice that the number of vortices relates to the values of $p_1$ and $p_2$ and in the non-Abelian case we notice that the value of $p_j$ mainly influences $\vert \Psi_{n,j} \vert ^2$ in the sense that  $\vert \Psi_{n,1} \vert ^2$ (red line) has $7$ vortices while $\vert \Psi_{n,2} \vert ^2$ (black line) has $3$ main vortices. We also notice that the general shape of the superfluid density (thick blue line) shows $3$ principal slopes.


A simple and general rule is observed for arbitrary values of the fluxes $\alpha_1$ and $\alpha_2$. In the Abelian case, for which $\alpha_1=\alpha_2=p/q$, with $p<q/2$, one naturally finds that $\vert \Psi_{n,1} \vert ^2 = \vert \Psi_{n,2} \vert ^2$ and exactly $p$ vortices are formed in the supercell of length $q$ (see Fig.\ref{plot}.a). In the non-Abelian case, for which $p_1 \ne p_2$, the ``color" densities differ $\vert \Psi_{n,1} \vert ^2 \ne \vert \Psi_{n,2} \vert ^2$ and the situation is complex: one finds $\tilde{p}$ vortices, with $\tilde{p}=max(p_1,p_2)$, but the overall shape of the total density $\vert \Psi _n \vert ^2$ is mostly influenced by $\hat{p}$, with $\hat{p}=min(p_1,p_2)$: in  Fig.\ref{plot}.b one can see $\tilde{p}=7$ local vortices and $\hat{p}=3$ general slopes. 

\begin{figure}
\begin{center}
\includegraphics[scale=1.1]{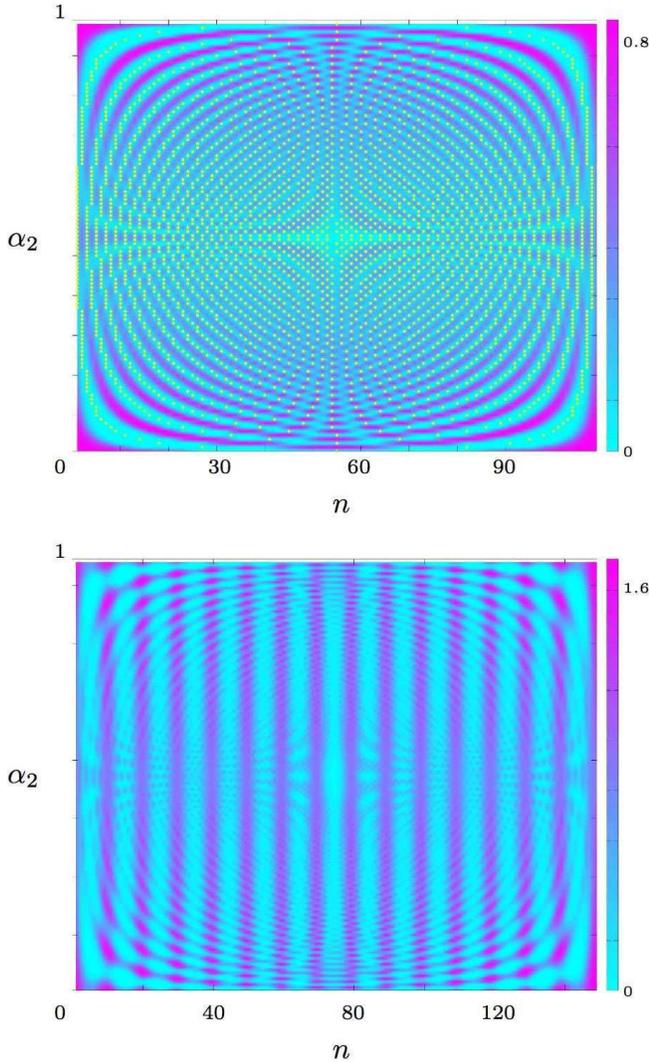}  \\
\end{center}
\caption{\label{density} Superfluid density  $\vert \Psi_{n} \vert ^2=\{ \Psi_n \vert \Psi_n \}$ as a fonction of $\alpha_2$ ($y$-axis) for $\mu=1.4$, $t_a=t_b=0.05$: (a) $\alpha
_1= \alpha_2$ with $q=109$ (yellow dots represent the vortices);  (b) $\alpha
_1=\frac{\sqrt{2}}{14}\approx \frac{15}{149}$ with $q=149$ and  $n=1,...,q$ ($x$-axis) }
 \end{figure}

This important result leads to a striking representation of the superfluid density as a fonction of the fluxes. In Fig.\ref{density}.a which illustrates the Abelian case, one observes the manner by which the number of vortices (yellow dots) evolves with the flux $\alpha=\alpha_1=\alpha_2$: there are exactly $p$ vortices within the supercell at $\alpha=\frac{p}{q}$, for $p<q/2$, and the superfluid density's general shape smoothly follows the rule through concentric circles (see the contour plot in Fig.\ref{density}.a). We notice that the figure is symmetric around the value $\alpha =0.5$ and is periodic in $\alpha$ with period one. In order to represent the non-Abelian case (Fig.\ref{density}.b) one set $\alpha_1=\frac{\sqrt{2}}{14}\approx \frac{15}{149}$ and $q=149$. In this case, $\alpha_1 \approx \frac{p_1}{q}$ with $p_1 = 15$ and we note that the result is robust for $\alpha_1$ irrational. The contour plot shows that the superfluid density is mainly affected by the fixed value $p_1 = 15$ which generates $15$ general slopes in the density for every value of $p_2>15$, but a pattern similar to the analoguous Abelian case is still observable in the background which correspond to the many $\tilde{p}=max(p_1,p_2)$ vortices formed as decribed here above. 
 
\begin{figure}
\begin{tabular}{lcr}
\hspace{-0.7cm} \includegraphics[scale=1.]{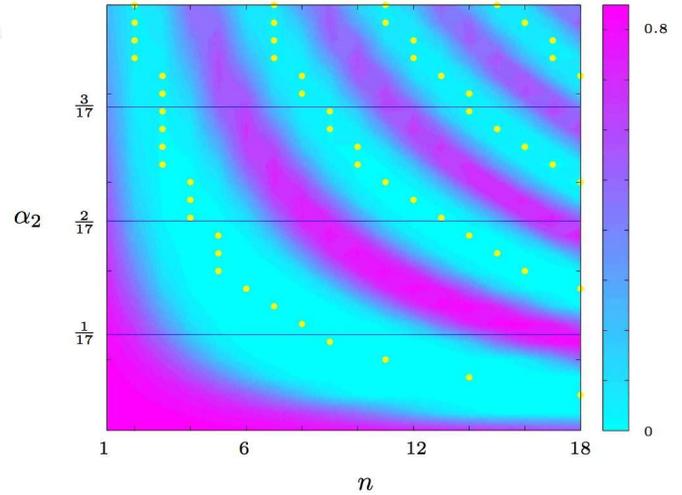}    
\end{tabular}
\caption{\label{dens} Superfluid density  $\vert \Psi_{n} \vert ^2=\{ \Psi_n \vert \Psi_n \}$ as a fonction of $\alpha_2$ ($y$-axis) for $\mu=1.4$, $t_a=t_b=0.05$, $\alpha
_1= \alpha_2=\frac{p_2}{q}$ for a supercell of length $q=109$ and $n=1,..,18$ ($x$-axis).  Yellow dots represent the vortices and the thin blue lines show the constant functions $\alpha_2=\frac{p}{Q}=\frac{1}{17}$, $\frac{2}{17}$ and $\frac{3}{17}$ where $1$, $2$ and $3$ vortices are respectively observed between $n=1$ and $n=Q=17$}
 \end{figure}

In the Abelian case, the superfluid density's representation in terms of the flux shows a particular property which we refer to as ``scale invariance". This property is summarized as follows: when the flux is rational, $\alpha=\frac{p}{q}$,  we have shown that $p$ vortices are formed whithin a $q\times 1$ supercell; but when considering a sample of $Q$ sites within the supercell, one finds exactly $p$ vortices at  $\alpha \approx \frac{p}{Q}$. This property is illustrated in Fig.\ref{dens} which shows the superfluid density computed in a $109\times 1$ supercell and represented as a function of the flux in a sample of length $Q=17$: as expected $p$ vortices are formed at values $\alpha=\frac{p}{17}$. An important consequence of this result is the fact that the number of vortices is independant of the supercell length's value $q$. Moreover, this property gives rise to the complex structure represented in Fig.\ref{density}a: the structure is formed in such a way that, between any interval $n=[1,Q]$ of arbitrary length $Q$, exactly $p$ dots are drawn at $\alpha=\frac{p}{Q}$. 

\begin{figure}
\begin{center}
\hspace{-0.8cm}
\includegraphics[scale=1]{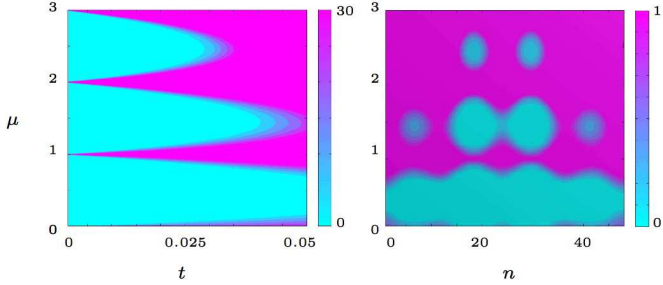}  
\end{center}
\caption{\label{phasedensity} Non-Abelian system for $p_1=1$, $p_2=4$ and $q=48$: (a) Phase diagram: total superfluid density  $\sum_n \vert \Psi_{n} \vert ^2$ as a fonction of the chemical potential $\mu$ ($y$-axis) and $t=t_a=t_b$ ($x$-axis); (b). Superfluid density  $\vert \Psi_{n} \vert ^2=\{ \Psi_n \vert \Psi_n \}$ as a fonction of the chemical potential $\mu$ ($y$-axis) with $t=t_a=t_b=0.05$ and $n=1,..,48$ ($x$-axis) }
\end{figure}

\begin{figure}
\begin{tabular}{lcr}
\hspace{-0.1cm} \includegraphics[scale=0.9]{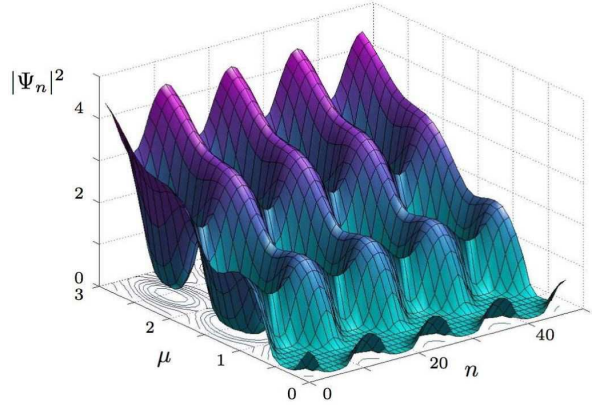}    
\end{tabular}
\caption{\label{mu} Superfluid density  $\vert \Psi_{n} \vert ^2=\{ \Psi _n \vert \Psi _n \}$ as a fonction of the chemical potential $\mu$ ($y$-axis) for $t_a=t_b=0.05$, $\alpha
_1= \alpha_2=\frac{4}{q}$, $q=48$ and $n=1,..,48$ ($x$-axis)}
 \end{figure}

The previous results presented here above are obtained for specific values of the hopping parameters $t_a$, $t_b$ and the chemical potential $\mu$. In Fig.\ref{phasedensity}.a we show the superfluid-insulator phase diagram in the $\mu -t$ plane for the non-Abelian case, with $t=t_a=t_b$, $p_1=1$, $p_2=4$ and $q=48$: three distinctive Mott lobes, inside which  $\sum_n \vert \Psi_{n} \vert ^2=0$, are clearly observed. In Fig.\ref{phasedensity}.b we represent the superfluid density while progressing vertically in the insulator-superfluid phase diagram at $t=0.05$: the superfluid density $\vert \Psi_{n} \vert ^2$ is vanishing in the whole supercell around $\mu=0.5$ which corresponds to the entrance of the first Mott lobe, then in the superfluid region, the vortex formation is important when approaching the Mott lobes at $\mu=1.5$ and $\mu=2.5$. The same phenomenon is observed in the Abelian case illustrated in Fig.\ref{mu} for $\alpha=\alpha
_1= \alpha_2=\frac{4}{48}$: four vortices are formed around $\mu=1.5$ (near the second Mott lobe) and $\mu=2.5$ (near the third Mott lobe) and vanish around $\mu=0.5$ (first Mott lobe).
 
 
In summary, the pattern formation resulting from gauge potentials created in optical lattices have been studied. The number of vortices formed in the superfluid Bose gase directly relates to the gauge potential values. In the Abelian case, this number results from a simple rule which leads to a salient figure that illustrates the superfluid density as a function of the effective magnetic flux. This phenomenon is observed in a wide region of the superfluid regime. As the vortices studied in this context have typical size $L \approx 5-10$ $a_0$ (with typical lattice constant $a_0=0.5-5$  $\mu m$, see Refs.\cite{Greiner1,Zwerger}), the patterns discussed in this work should be directly observed in laboratories by optical methods as suggested by Wu et al. in Ref. \cite{Wu}.  

\acknowledgments
N. G. thanks Pierre Gaspard, Pierre de Buyl, Serge Goldman and Rita Matos Alves for useful discussions and the F.~R.~I.~A. $-$ F.~N.~R.~S. for financial
support. Special thanks to Nassiba Tabti, Vincent Wens, Jean-Sabin Mc Ewen, Ella Jamsin, Jon Demayer, Julie Delvax, Yassin Chaffi, Aina Astudillo and David Melviez.

\end{document}